\documentclass[12pt,aps]{revtex4}

\usepackage{graphicx}
\begin{document}
\title{Modeling and Simulation of a Microstrip-SQUID Amplifier}
\author{G.P. Berman$^a$,  O.O. Chumak$^b$,
D.I. Kamenev$^{a}$, D. Kinion$^{c}$, and V.I. Tsifrinovich$^{d}$}
\affiliation{$^a$Theoretical Division, Los Alamos National
Laboratory, Los Alamos, NM 87545, USA} \affiliation{$^b$Institute of
Physics of the National Academy of Sciences, Pr. Nauki 46, Kiev-28,
MSP 03028, Ukraine} \affiliation{$^c$Lawrence Livermore National Laboratory,
Livermore, CA 94551, USA} 
\affiliation{$^d$Department of Applied Physics,
Polytechnic Institute of NYU, 6 MetroTech Center, Brooklyn, NY
11201, USA}

\begin{abstract}
Using a simple lumped-circuit model, we numerically study the dependence of the voltage 
gain and noise on the amplifier's parameters. Linear, quasi-linear, and
nonlinear regimes are studied. We have shown that the voltage gain 
of the amplifier cannot exceed a characteristic critical value, which 
decreases with the increase of the input power. We have also shown 
that the spectrum of the voltage gain depends significantly on the level 
of the Johnson noise generated by the SQUID resistors. 
\end{abstract}

\maketitle 
\section{Introduction}
\begin{figure}
\centerline{\includegraphics[width=13cm,height=7cm,angle=0,clip=]{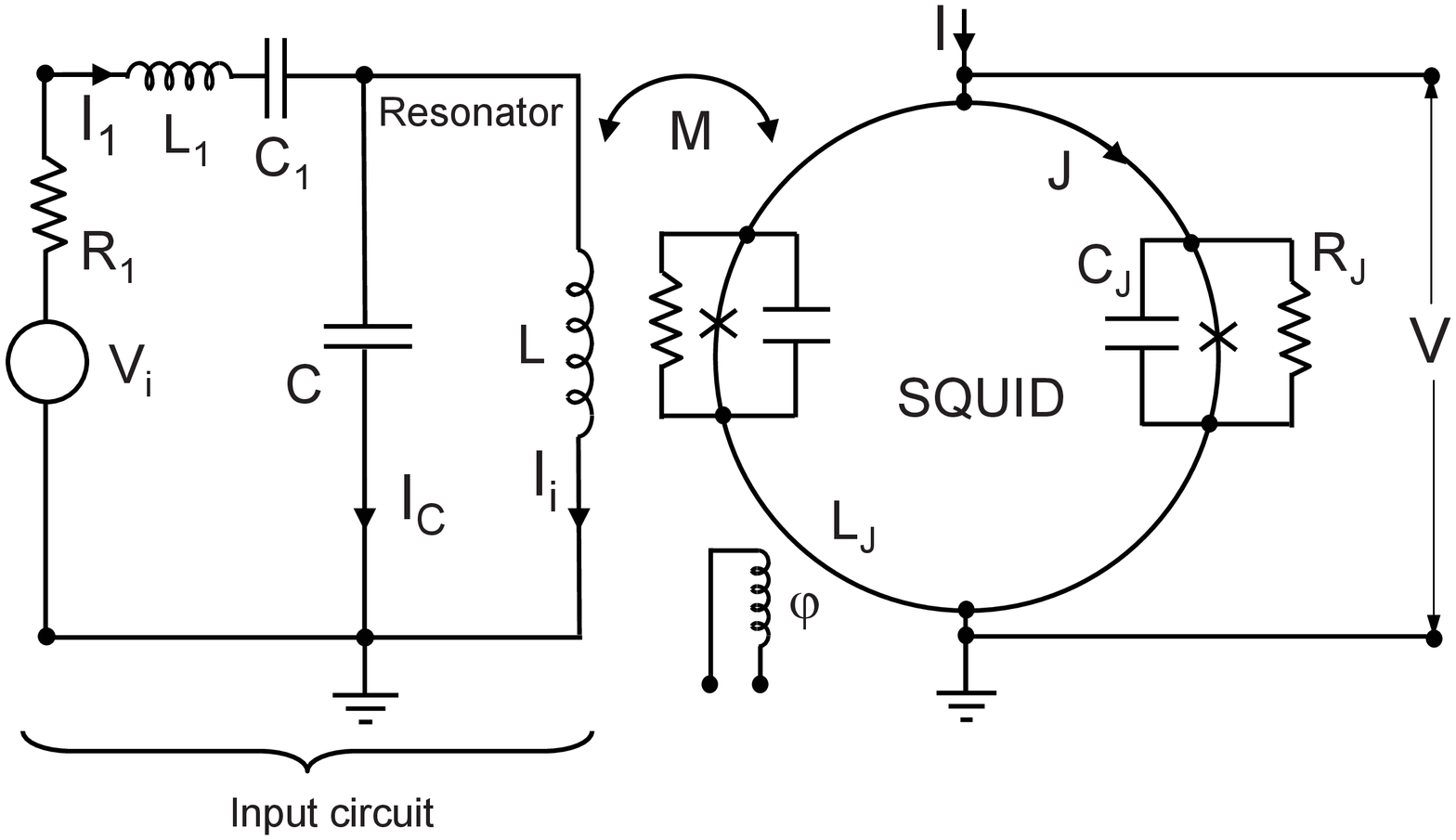}}
\vspace{-6mm} \caption{The equivalent scheme of the microstrip-SQUID
amplifier. $V_i$ is the amplitude of the input voltage and $R_1$ is
the resistance of the voltage source.  $L_1$ describes the stray
inductance, and (if necessary) the pick-up coil inductance, $C_1$ is the
coupling capacitance, $L$ is the inductance of the input coil for
the SQUID, $I$ and $\varphi$ are, respectively, the bias current and
the flux for the SQUID, $M$ is the mutual inductance between the
input coil, $L$, and the SQUID loop. $L_J$ is the inductance of the
 SQUID loop, $I_1$, $I_C$, and $I_i$ are the currents in the input
part of the circuit, and $J$ is the current circulating around the
SQUID loop, $C_J$ and $R_J$ are, respectively, the capacitance and
resistance of each Josephson junction.} \label{reffig:1}
\end{figure}

A microstrip-SQUID (superconducting quantum interference device)
amplifier (MSA) has been designed as a low noise radiofrequency
amplifier, which is able to operate above 100 MHz~\cite{1}. The MSA
has been studied theoretically in many
publications~\cite{1,referee1,referee2,2,3,4,5,6,7}. However, a consistent theoretical
model of MSA has not yet been developed. The circuit diagram of
lumped model of MSA is presented in Fig.~\ref{reffig:1}. This MSA
consists of a linear input circuit coupled to the direct current
(dc) SQUID via the mutual inductance, $M$. Note that the isolated dc
SQUID is a nonlinear circuit, while the isolated microstrip is a
linear circuit. The total system consisting of the SQUID and the
input circuit is a nonlinear one. 
Consequently, it remains no trivial task to predict the performance 
of the MSA during the design process. Therefore analytical investigation
and extensive numerical modeling, simulation, and optimization of the MSA are required before
creating the device. 
Generally, the solutions for the MSA model must be obtained by solving 
analytically or numerically the system of nonlinear ordinary differential equations written for the
SQUID coupled to the input circuit.  The dynamics of a bare SQUID
was investigated numerically in~\cite{tesche1,tesche2,tesche3}.

The objective of our paper is to study numerically the dynamics of
exact nonlinear equations describing the MSA, and to compute the voltage gain,
$G(f)=|V(f)/V_i|$, of the MSA, where $V(f)$ is a Fourier harmonic of
the output voltage on the SQUID, and $V_i$ is the amplitude of the
input voltage on the microstrip at the same frequency, $f$. We
analyze numerically both linear and nonlinear regimes of
amplification. A linear regime means that the following linear
dependence exists: $|V(f)|=G(f) |V_i|$, where the gain $G(f)$ is
independent of the $V_i$.  We also simulate the output spectral
density of voltage Johnson noise, originated in shunting resistors
of the SQUID and the resistor, $R_1$, in the input circuit, and 
calculate the noise temperature.

\section{Input circuit}
Consider the isolated linear input circuit ($M=0$). The forward
impedance of the input circuit is

\begin{equation}
Z_i={V_i\over I_i}=\left(2\pi if L_1+{1\over 2\pi if C_1} +R_1\right)
\left(1-(2\pi f)^2LC\right)+2\pi if L.
\end{equation}

\begin{figure}
\centerline{\includegraphics[width=11cm,height=9cm,angle=0,clip=]{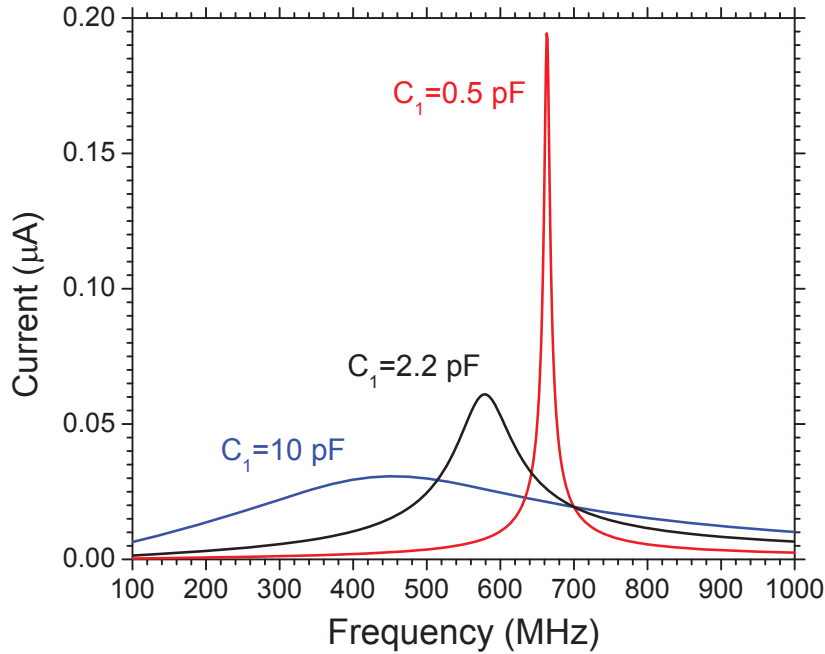}}
\vspace{-2mm} \caption{Current, $I_i$, in the input coil for an
isolated input circuit ($M=0$), and for three values of the coupling
capacitance, $C_1$. $R_1=50$ Ohm, $V_i=1$ $\mu$V, $L_1=1$ nH,
$C=4.4$ pF. $f_0=1/(2\pi\sqrt{LC})=700$ MHz is the resonant
frequency of the input resonator.} \label{reffig:ic}
\end{figure}

In Fig.~\ref{reffig:ic} we plot the amplitude of the current, $I_i$, in the
input coil with inductance $L$, when $M=0$. One can
see from Fig.~\ref{reffig:ic} that as $C_1$ decreases, the maximum
shifts to higher frequencies, and the width of the peak decreases.
One can use this latter property to create a narrow-bandwidth
amplifier. The frequency corresponding to the maximum is always less
than the resonant frequency, $f_0$, of the input resonator.

\section{Equations of motion}
The differential equations of motion for the SQUID are:~\cite{2}
\begin{eqnarray}
\label{squid}
& &\varphi_0C_J\ddot \delta_1+{\varphi_0\over R_J}\dot\delta_1=
{I\over 2}-J-I_0\sin\delta_1+I_{n1},\nonumber\\
& &\varphi_0C_J\ddot \delta_2+{\varphi_0\over R_J}\dot\delta_2=
{I\over 2}+J-I_0\sin\delta_2+I_{n2},\\
& &\varphi_0(\delta_1-\delta_2)=\varphi+L_JJ+MI_i\nonumber.
\end{eqnarray}
Here the dot above $\delta_1$ and $\delta_2$ indicates time differentiation;
$\varphi_0=\hbar/(2e)$ is the reduced flux quantum;
$\hbar$ is Planck's constant; $e$ is the electron charge;
$\delta_1$ and $\delta_2$ are the phase differences in the Josephson
junctions in the SQUID; $I_0$ is the Josephson junction critical
current; $I_{n1}$ and $I_{n2}$ describe the noise current (Johnson
noise) originating in the shunt resistors, $R_J$. The
output voltage, $V$, can be expressed in terms of the Josephson
junction phase differences
$$
V={\varphi_0\over 2}\left(\dot\delta_1+\dot\delta_2\right).
$$
In order to determine $I_i$ in the third equation in (\ref{squid}),
we have to add the differential equations for the input circuit. The
total system of eight first-order differential equations can be
written in the following form:
\begin{eqnarray}
\label{amp}
& &\dot x_1=p_1,\nonumber\\
& &\dot p_1=
{1\over C_J}\left(-{1\over R_J}p_1+I-2\sin{x_1\over 2\varphi_0}\cos{x_2\over 2\varphi_0}+
I_{n1}+I_{n2}\right),
\nonumber\\
& &\dot x_2=p_2,\nonumber\\
& &\dot p_2={1\over C_J}
\left[-{1\over R_J}p_2-{1\over L_J}(x_2-\varphi-MI_i)-2\sin{x_2\over 2\varphi_0}
\cos{x_1\over 2\varphi_0}+I_{n1}-I_{n2}
\right],\\
& &\dot Q_1=I_1,\nonumber\\
& &\dot I_1={1\over L_1}\left[V_i\cos(2\pi f t)+V_{n}
-{Q_1\over C_1}-I_1R_1-{Q_C\over C}\right],
\nonumber \\
& &\dot Q_C=I_1-I_i,\nonumber\\
& &\dot I_i={1\over L\alpha}\left({Q_C\over C}-{M\over L_J}p_2\right)\nonumber.
\end{eqnarray}
Here
$$
x_1=\varphi_0(\delta_1+\delta_2);\qquad x_2=\varphi_0(\delta_1-\delta_2);\qquad
\alpha=1-{M^2\over LL_J};
$$
$Q_1$ is the charge on the capacitor $C_1$; $Q_C$ is the charge on the capacitor $C$; 
$f$ is the frequency of the external voltage; and $V_{n}=I_nR_1$ is the noise voltage on the 
the resistor $R_1$; $I_n$ is the noise current through $R_1$.

The input circuit  is coupled to the SQUID through the term,
$MI_i/(C_JL_J)$, in the fourth equation in~(\ref{amp}), and the SQUID
is coupled to the input circuit by the effective coupling constant,
$\gamma=M/(LL_J\alpha)$, in the last equation in~(\ref{amp}). Since
the effective coupling constant, $\gamma$, is proportional to
$1/\alpha$, the effective coupling can be increased by decreasing
$\alpha$.

\begin{figure}
\centerline{\includegraphics[width=8cm,height=6cm,angle=0,clip=]{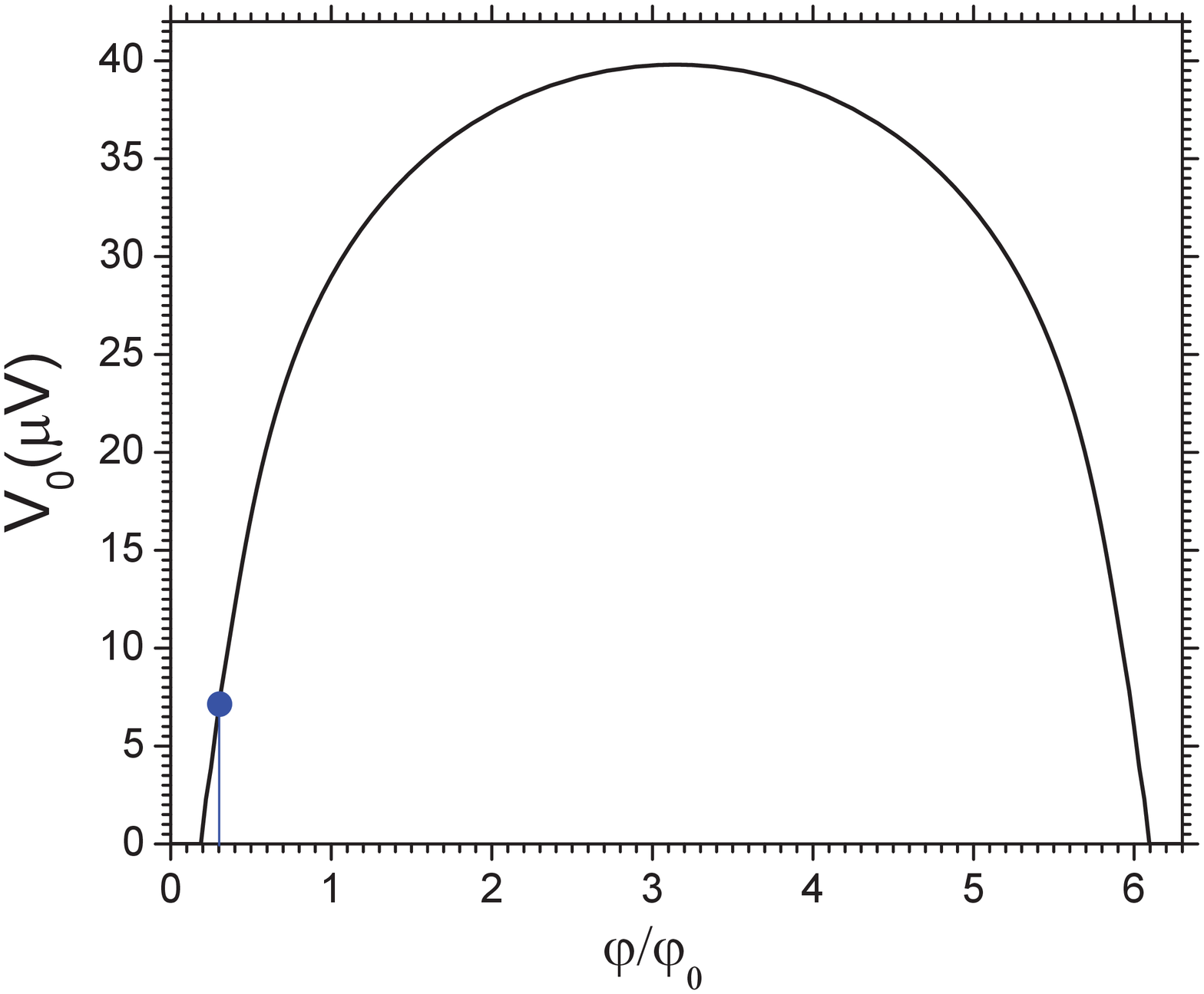}\vspace{5mm}
\includegraphics[width=8cm,height=6cm,angle=0,clip=]{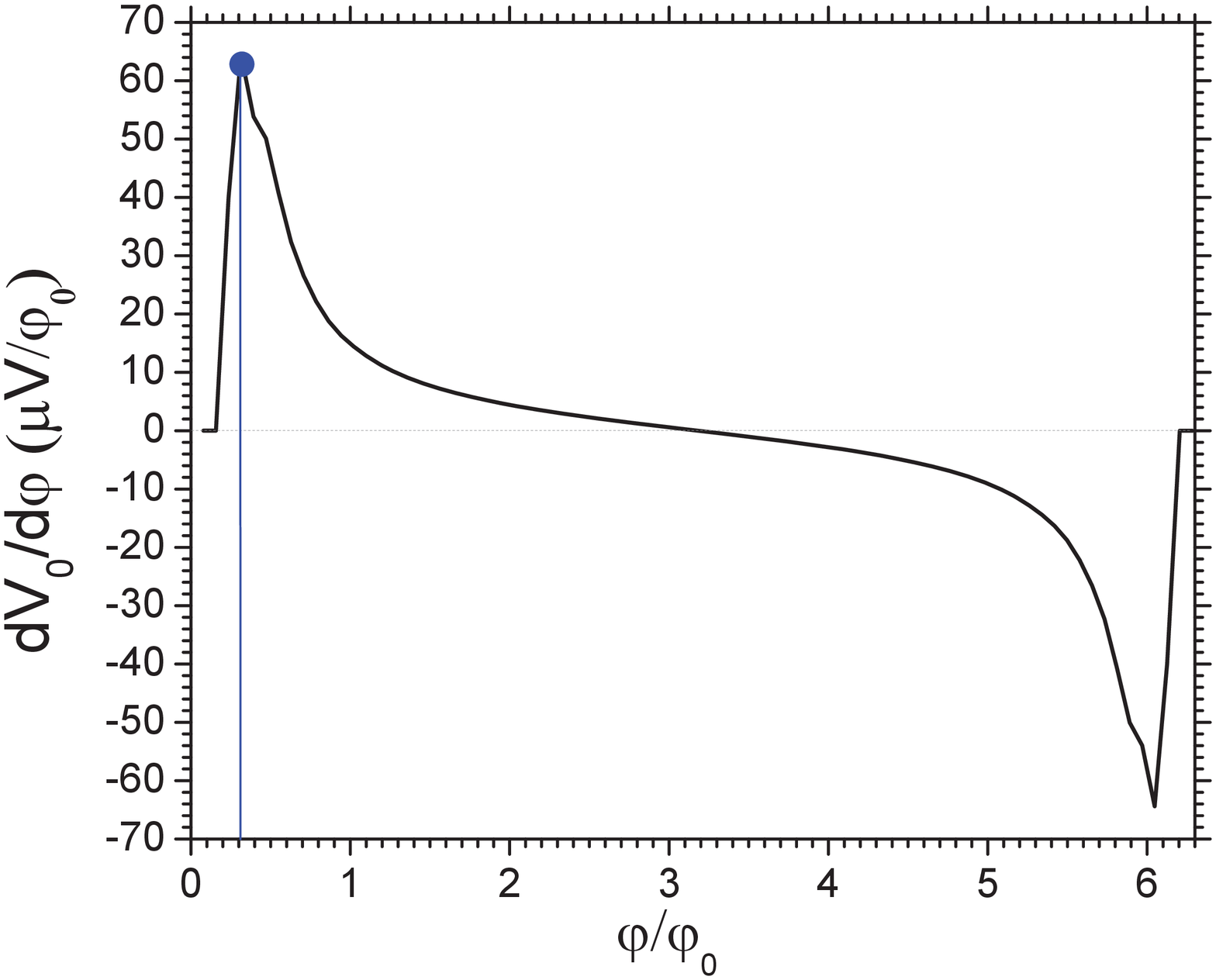}}
\vspace{-6mm} \caption{Average over time of the output voltage,
$V_0$, (left) and its derivative (transfer function) (right). The
working point for $\varphi/\varphi_0=0.3$ is marked by a filled blue
circle. $C_J=0.2$ pF, $L_J=0.45$ nH, $R_J=20$ Ohm, $I=1.99I_0$,
$I_0=2$ $\mu$A, $\alpha=0.001$, $C_1=0.5$ pF, $V_i=0$,
$I_{n1}=I_{n2}=0$. The other parameters are the same as in
Fig.~\ref{reffig:ic}.} \label{reffig:phi}
\end{figure}

\section{Voltage gain}
First we choose the optimal working point of our amplifier which is
defined by the value of $\varphi$, provided the other parameters are
given. In Fig.~\ref{reffig:phi} the time average, $V_0$, of the output voltage
and its derivative (transfer function) are plotted as a function of $\varphi$.
The maximum of the transfer function occurs in the vicinity of
$\varphi/\varphi_0=0.3$ which we choose as our working point.

\begin{figure}
\centerline{\hspace{-5mm}\includegraphics[width=8cm,height=6cm,angle=0,clip=]{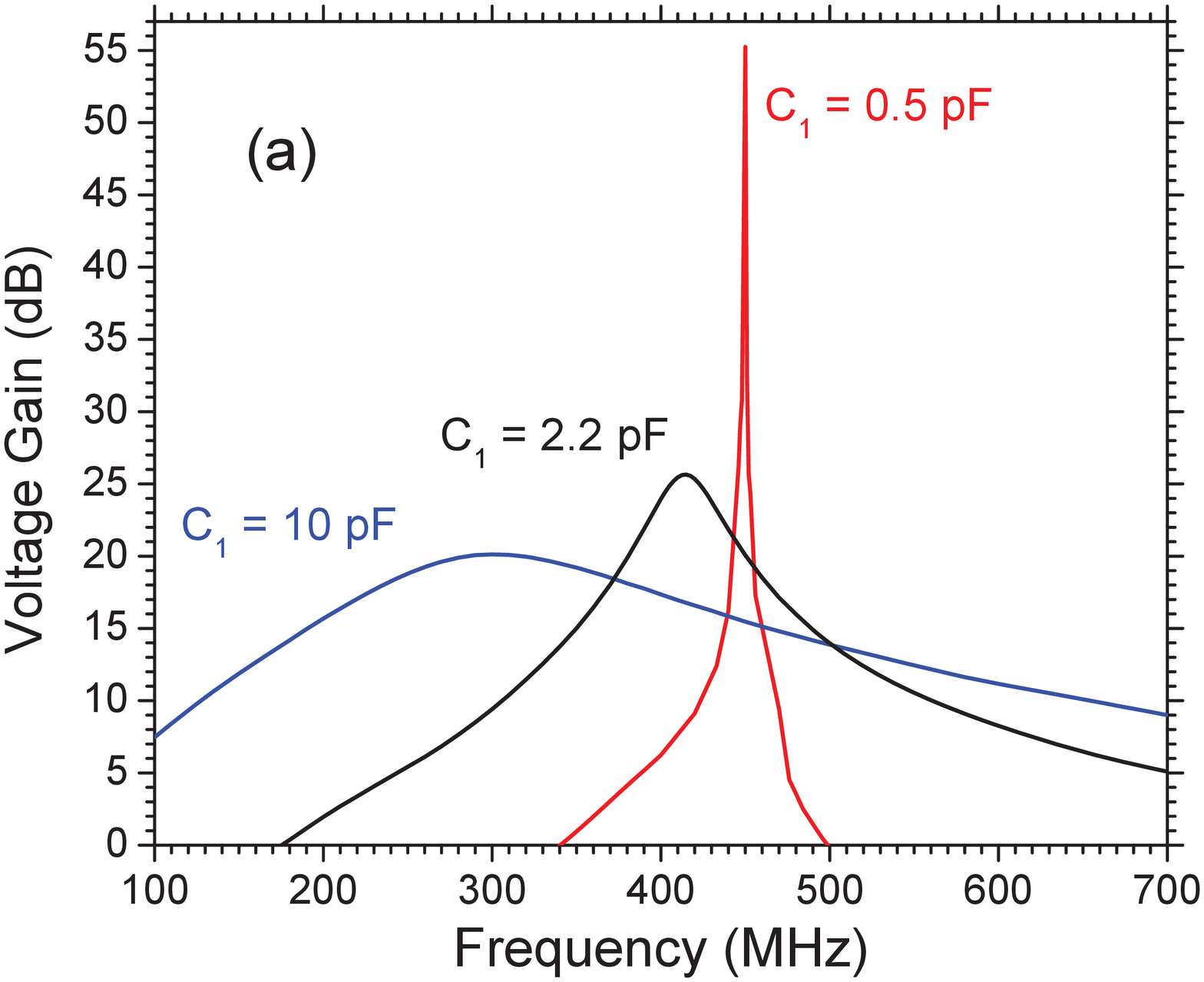}
\hspace{-2mm}\includegraphics[width=8cm,height=6cm,angle=0,clip=]{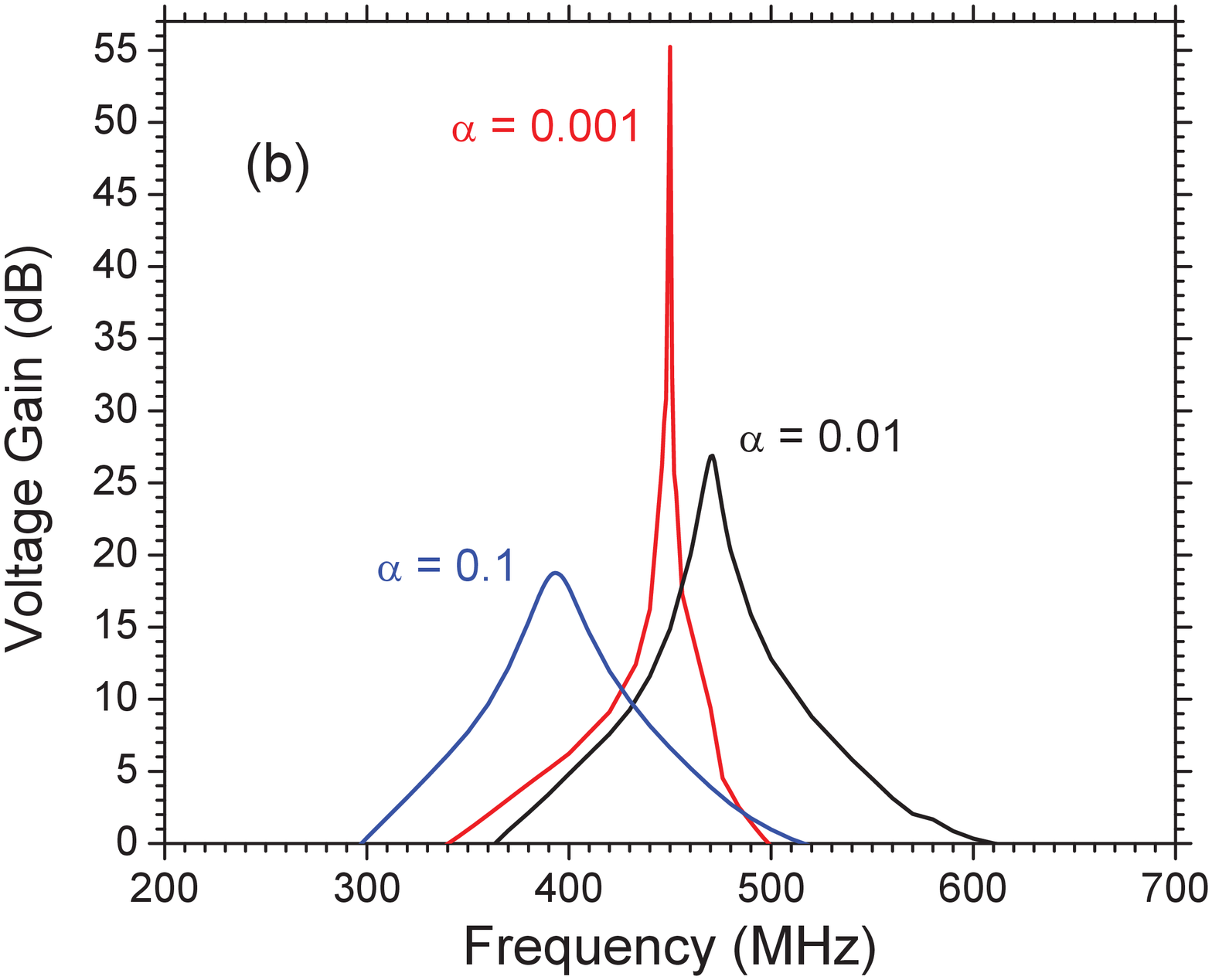}}
\vspace{-3mm} \caption{(a) Voltage gain of the amplifier for three
values of the coupling capacitance, $C_1$, and for $\alpha=0.001$.
(b) Voltage gain for three values of $\alpha$, and for $C_1=0.5$ pF.
$V_i=0.001$ $\mu$V in both (a) and (b).
The other parameters are the same as in Figs.~\ref{reffig:ic} and
\ref{reffig:phi}.} \label{reffig:gain}
\end{figure}

In Fig.~\ref{reffig:gain}(a) we plot the voltage gain for three
values of the coupling capacitance, $C_1$. By comparison of
Fig.~\ref{reffig:gain}(a) with Fig.~\ref{reffig:ic}, one can
conclude that the amplifier gain is mostly defined by the parameters
of the input circuit. The gain of 55 dB for $C_1=0.5$ pF is the
combined result of amplification by the input circuit and by the
SQUID due to the strong interaction between them when $\alpha$ is
small and positive. Note that the dimensionless parameter, $\alpha$, contains both the
parameters of  the SQUID ($L_J$) and the input circuit ($L$), as
well as the coupling inductance, $M$. In Fig.~\ref{reffig:gain}(b)
we plot the gain for three values of $\alpha$: $\alpha=0.001$, 0.01,
and 0.1. The gain decreases from 55 dB to 27 dB as $\alpha$
increases from 0.001 to 0.01. The value of coupling inductance, $M$,
changes, respectively, from 2.2982 nH to 2.2878 nH, that is, by only
0.44 percent. Therefore, the possibility of obtaining the large
gain is limited to a very small region of the parameters. For this
purpose it is desirable to have a tunable coupling inductance, $M$,
or a tunable input circuit inductance, $L$, or a tunable SQUID
inductance, $L_J$.

Consider the situation in which $\alpha$ is negative, that is $M^2>LL_J$. 
In order to understand the dynamics in this regime, we
differentiate the last equation in~(\ref{amp}) and use the 4th and
7th equations for $\dot Q_C$ and $\dot p_2$. We obtain an equation
for the oscillations of the input current, $I_i$, with an external
force and with the eigenfrequency, $\omega_0$, where
$$
\omega_0^2={1\over \alpha}
\left({1\over LC}+{M^2\over L_J^2LC_J}\right).
$$
Negative $\alpha$ corresponds to negative real part of the input 
impedance~\cite{negative2,negative3}. This regime 
can drive the resonator into instability~\cite{negative1}.

\begin{figure}
\centerline{\includegraphics[width=13cm,height=10cm,angle=0,clip=]{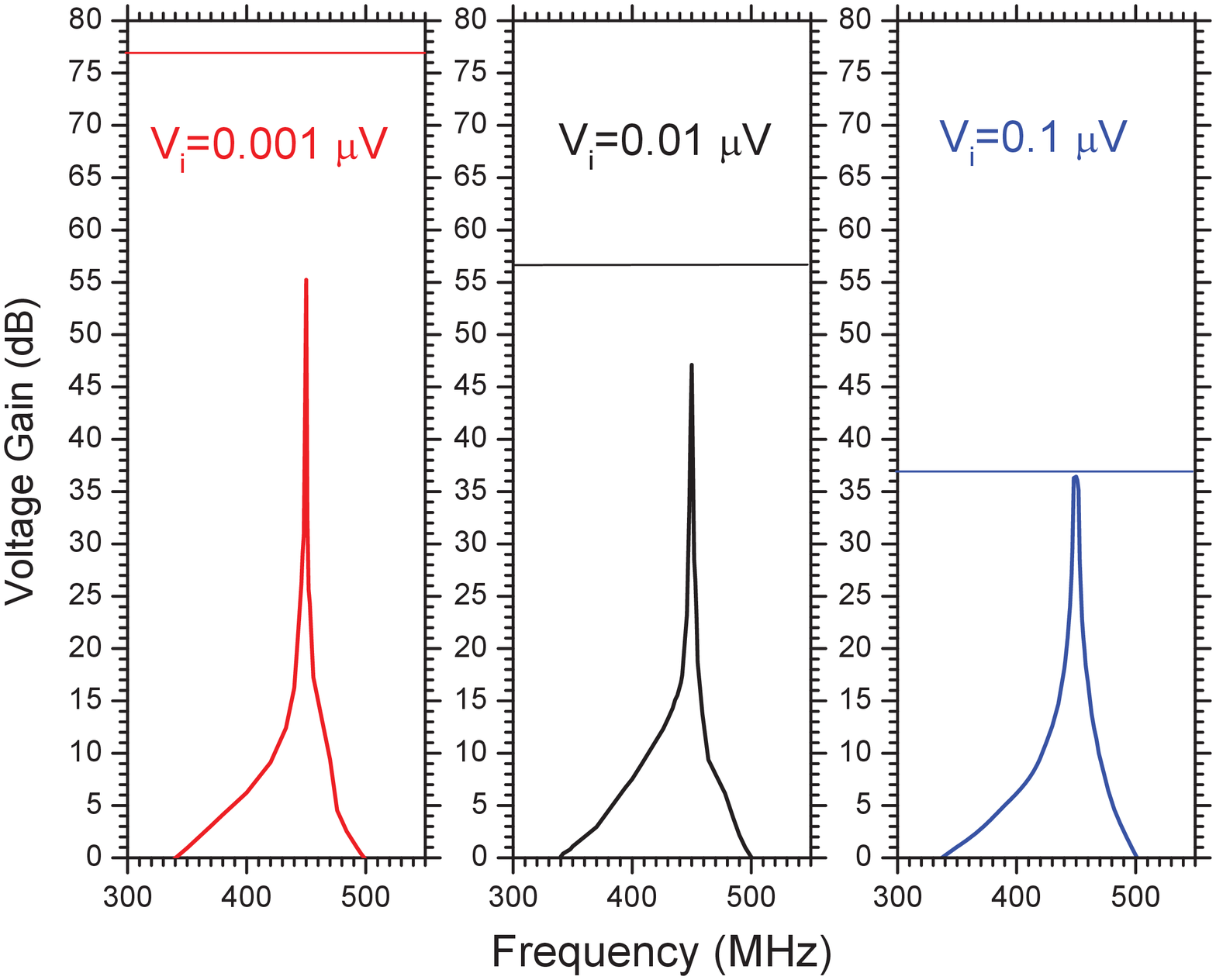}}
\vspace{-3mm} \caption{Voltage gain for three values of input
voltage amplitude $V_i$. $\alpha=0.001$, $C_1=0.5$ pF, the other
parameters are the same as in Figs.~\ref{reffig:ic} and
\ref{reffig:phi}.} \label{reffig:Vi}
\end{figure}

\section{Nonlinearity}
\begin{figure}
\centerline{\includegraphics[width=8cm,height=6cm,angle=0,clip=]{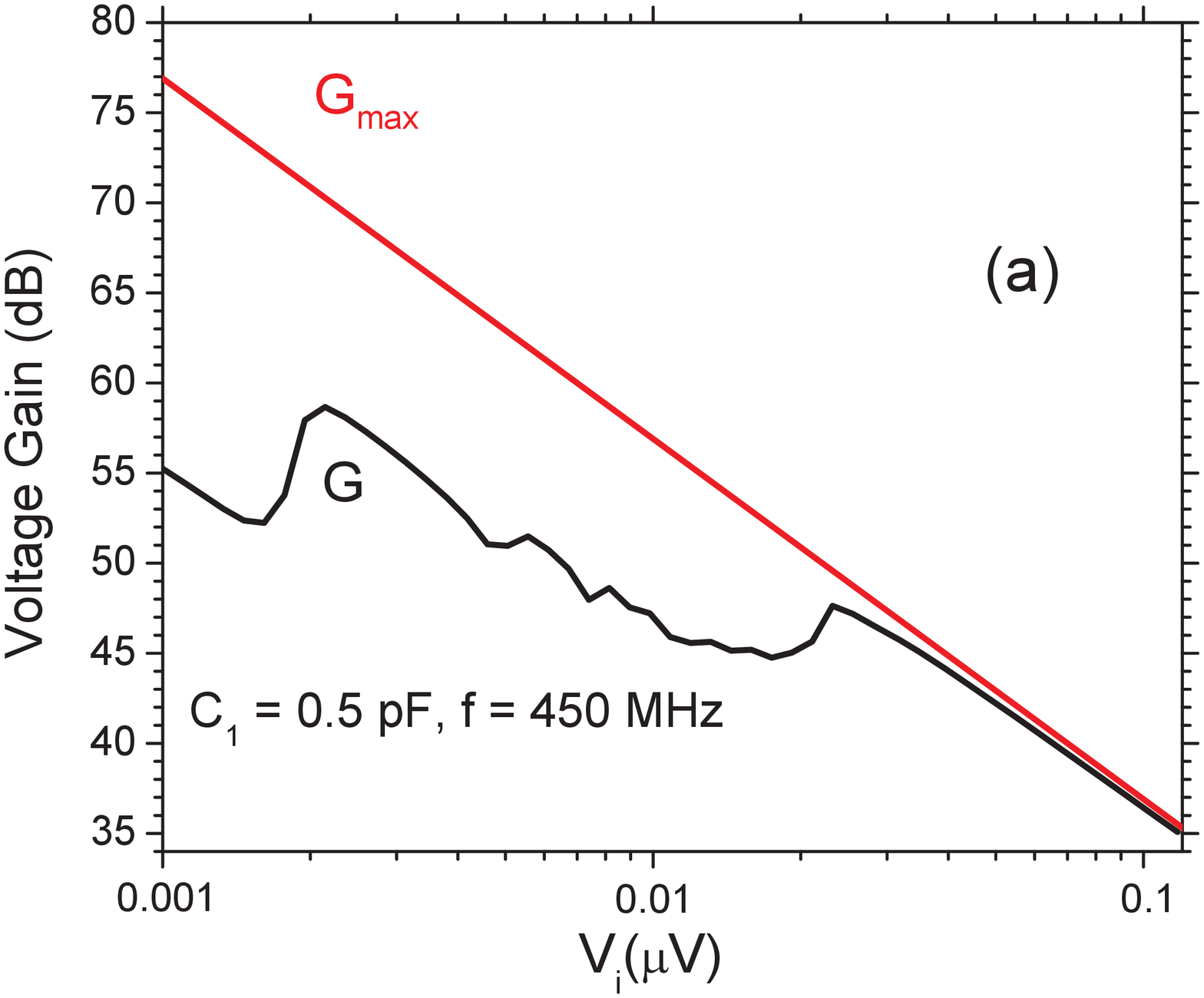}
\hspace{-3mm}\includegraphics[width=8cm,height=6cm,angle=0,clip=]{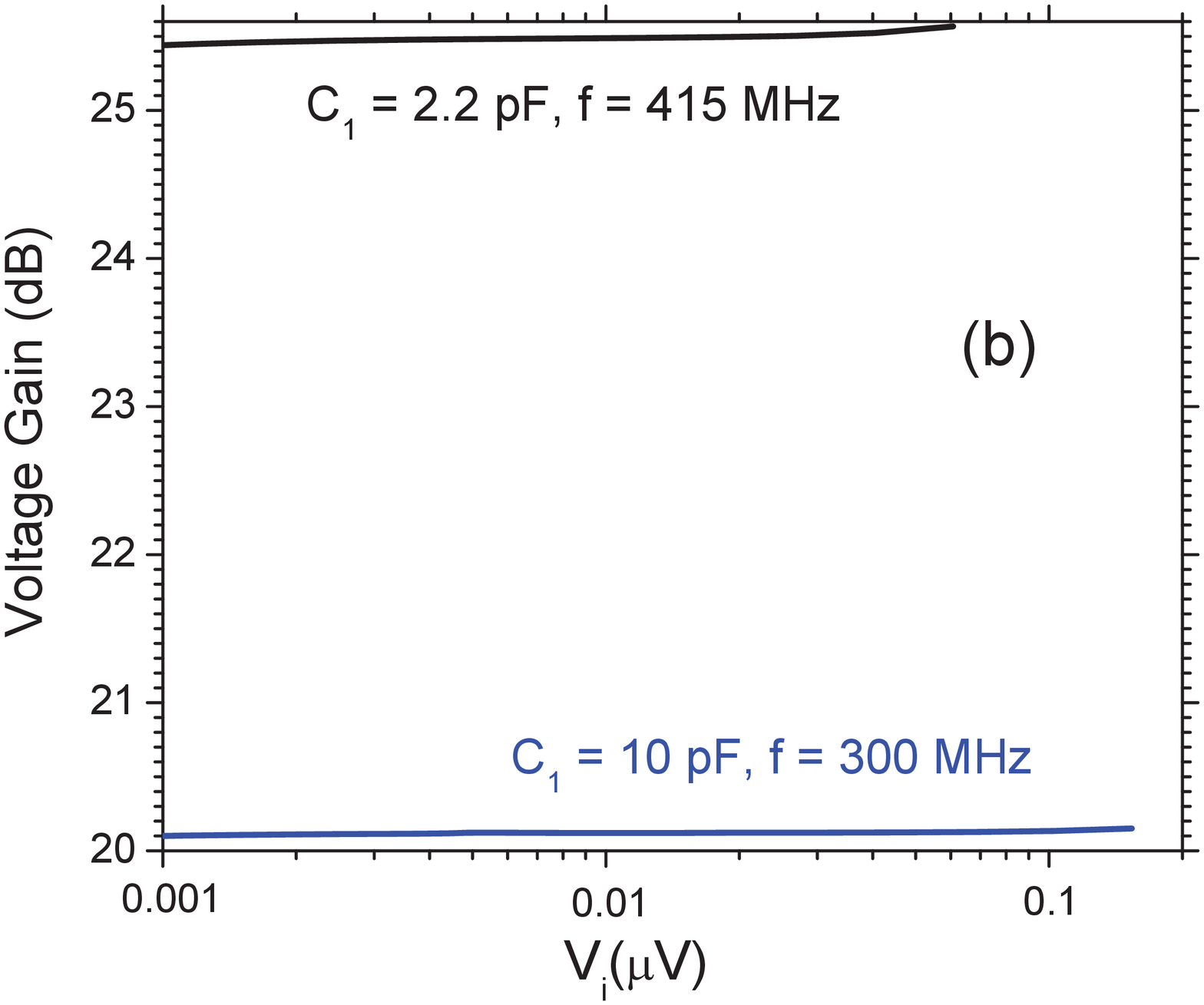}}
\vspace{-3mm} \caption{(a) Voltage gain, $G$, and maximum gain,
$G_{max}$, as a function of the input voltage amplitude, $V_i$. In
maximum $G=58.7$ dB (black curve). (b) Voltage gain, $G$, in linear regime 
as a function of the input voltage amplitude, $V_i$. $\alpha=0.001$
and the other parameters are the same as in
Figs.~\ref{reffig:ic} and \ref{reffig:phi}.} \label{reffig:maxV}
\end{figure}

When the input voltage amplitude or gain becomes sufficiently large,
the nonlinear effects in the SQUID become important. In the
nonlinear regime, the nonlinear effects decrease the output voltage
of the SQUID, thus decreasing the gain of the amplifier in
comparison with the linear regime. It is reasonable to assume that
the MSA is in the nonlinear regime when the output voltage becomes
comparable with the SQUID's own average output voltage, $V_0$. It is
convenient to define the maximum gain
\begin{equation}
\label{Gmax}
G_{max}=\left|{V_0\over V_i}\right|.
\end{equation}
When the amplitude, $|V(f)|$, of the amplified output Fourier
harmonic approaches $|V_0|$, the gain should decrease due to
nonlinear effects in the SQUID. If, for example, the amplitude of
the input signal is $V_i=1$ $\mu$V and $V_0=7$ $\mu$V (see left side
in Fig.~\ref{reffig:phi} for $\varphi/\varphi_0=0.3$), according to
Eq.~(\ref{Gmax}) the gain is limited by the value $G_{max}=7$. In
order to obtain a gain of $G=55$ dB in Figs.~\ref{reffig:gain}(a)
and (b), we set the amplitude of the input signal to be $V_i=0.001$
$\mu$V. For this input signal expressed in decibels 
$\left[G_{max}(dB)\rightarrow 20\log_{10}G_{max}\right]$, we have
$G_{max}=76.9$ dB, so that
$G(f)<G_{max}$. For $V_i=0.01$ $\mu$V, we have $G_{max}=56.9$ dB,
and for $V_i=0.1$ $\mu$V one obtains $G_{max}=36.9$ dB. In
Fig.~\ref{reffig:Vi} we plot the gain, $G(f)$, for three values of
$V_i$. One can see from the figure that $G(f)$ is less than
$G_{max}$ for all frequencies, $f$. (The values of $G_{max}$ for
each $V_i$ are indicated in the figures by the horizontal lines.) In
Fig.~\ref{reffig:maxV}(a) we plot $G$ as a function of $V_i$. 
As follows from the figure, the condition $G<G_{max}$
is also satisfied for all input voltage amplitudes, $V_i$. As the
amplitude, $V_i$, of the input signal increases, both $G_{max}$ in
Eq.~(\ref{Gmax}) and the gain, $G$, decrease. This nonlinear effect
cannot be obtained from a linearized theory, like that based on an
effective impedance of the amplifier~\cite{martinis}.
In Fig.~\ref{reffig:maxV}(b) the amplifier is in the linear regime
for $C_1=2.2$ pF and $C_1=10$ pF because the gain is sufficiently small, $G\ll G_{max}$. 

\section{Noise}
\begin{figure}
\centerline{\includegraphics[width=10cm,height=8cm,angle=0,clip=]{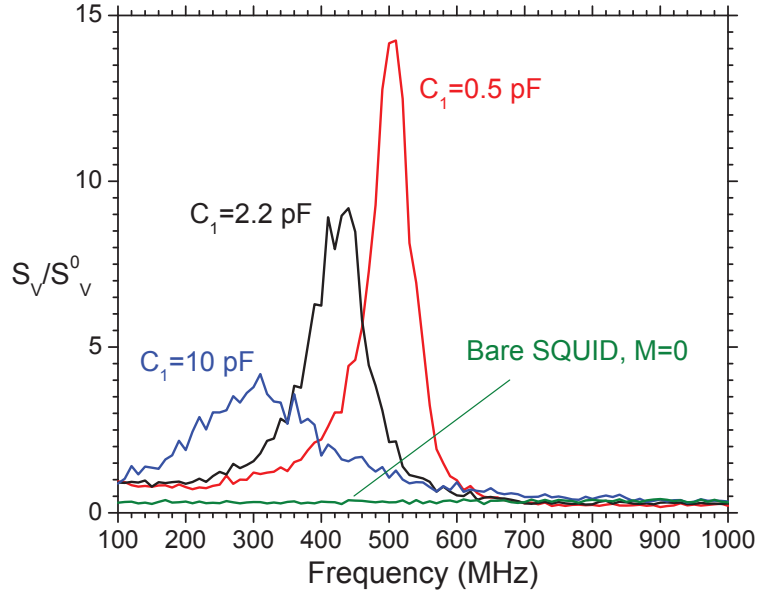}}
\vspace{-3mm} \caption{Output voltage spectral density $S_V/S_V^0$,
where $S_V^0=\varphi_0 I_0 R_J$, as a function of frequency, $f$,
for three values of $C_1$, $\alpha=0.001$. For comparison, we also
include the plot of $S_V/S_V^0$ for bare SQUID  (when $M=0$) with
white noise spectral density (because the frequency, $f$, is much
smaller than the Josephson frequency). The time of integration is
$t_i=10^4t_0$, where $t_0=\varphi_0/(I_0R_J)=8.23\times 10^{-12}$ s;
the number of realizations is $N=100$; $T=100$ mK; $\Delta
t=0.1t_0$;
 $V_i=0$;
the other parameters are the same as in
Figs.~\ref{reffig:ic} and \ref{reffig:phi}.}
\label{reffig:Sv}
\end{figure}

We assume that the Johnson noise voltages across the resistors
dominate all other sources of noise in the amplifier. White Gaussian
noise in each of three resistors was modeled by a train of
rectangular pulses following each other without interruption. The
current amplitude, $I_n$, of each pulse is random with zero average
and the following variance:
$$
\langle I_n^2\rangle = 2{k_BT\over R\Delta t}.
$$
Here the brackets indicate an average over different realizations;
$k_B$ is the Botzmann constant; $T$ is the temperature of the MSA;
$\Delta t$ is the duration of each pulse, which is constant in our
simulations; $R$ is the corresponding resistance: $R=R_J$ for the
shunting resistors in the SQUID and $R=R_1$ for the resistor $R_1$
in the input circuit. In Fig.~\ref{reffig:Sv} we plot the output
voltage spectral density, $S_V(f)/S_V^0$ (where $S_V^0=\varphi_0 I_0 R_J$),
when no input voltage is
applied to the input circuit, $V_i=0$. $S_V$ is defined as:
$$
S_V=\lim_{t_i\rightarrow \infty} {2\over
t_i}\langle|V(f)|^2\rangle,\qquad V(f)=\int_0^{t_i} V(t)e^{2\pi if
t}dt,
$$
where $t_i$ is the time of integration of the output signal and the
angular brackets, $\langle\rangle$, indicate an average over
different realizations.

\begin{figure}
\centerline{\hspace{-5mm}\includegraphics[width=8cm,height=6cm,angle=0,clip=]{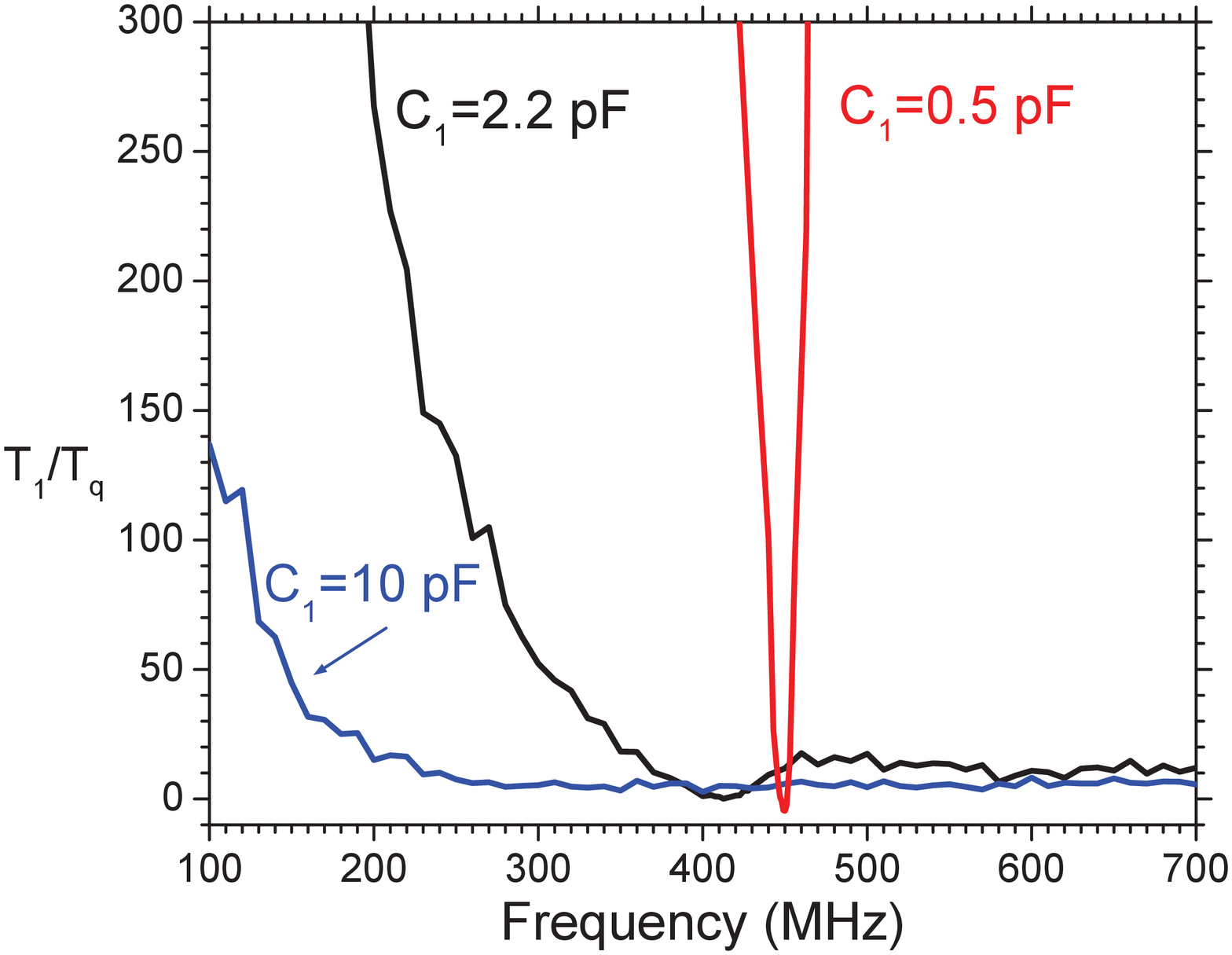}
\hspace{-3mm}\includegraphics[width=8cm,height=6cm,angle=0,clip=]{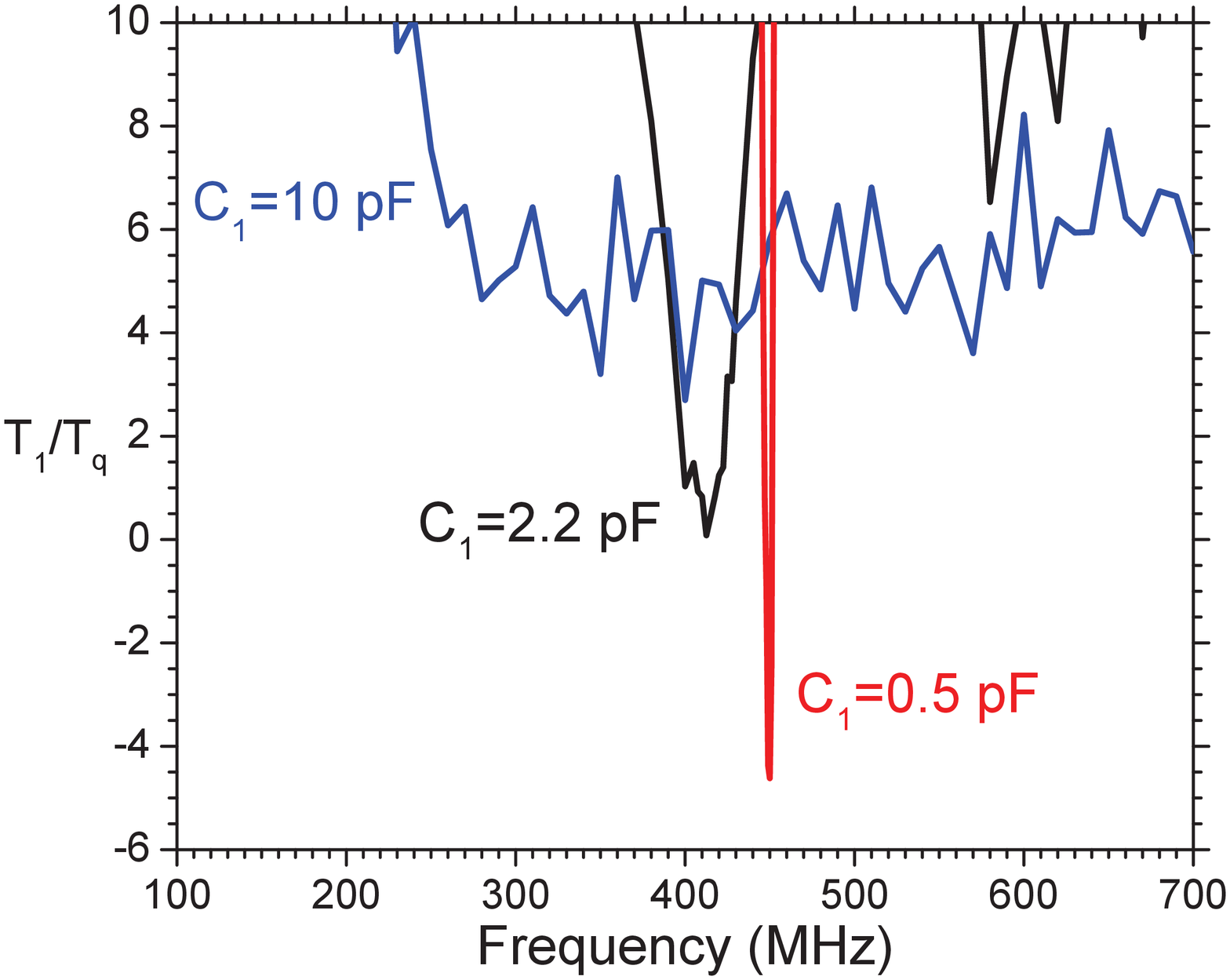}}
\vspace{-3mm}
\caption{Noise temperature $T_1/T_q$ for three values of $C_1$ in two different scales.
The other parameters are the
same as in Fig.~\ref{reffig:gain}(a) and Fig.~\ref{reffig:Sv}.}
\label{reffig:TC}
\end{figure}

One can use the voltage spectral density, $S_V(f)$, to calculate
the noise temperature of the amplifier, $T_1$, using the following equation:
$$
4k_B(T_1+T)R_1\left|G(f)\right|^2=S_V(f).
$$
In Fig.~\ref{reffig:TC} we plot $T_1/T_q$,  as a function of
frequency, $f$, for two different scales, where $T_q=hf/k_B$ is the
quantum temperature. We use the noise spectral density, $S_V(f)$, from
Fig.~\ref{reffig:Sv} and the gain from~Fig.~\ref{reffig:gain}(a). At
the minimum, for $C_1=0.5$~pF (red line) the noise temperature is
negative.

\begin{figure}
\centerline{\hspace{-5mm}\includegraphics[width=10cm,height=8cm,angle=0,clip=]{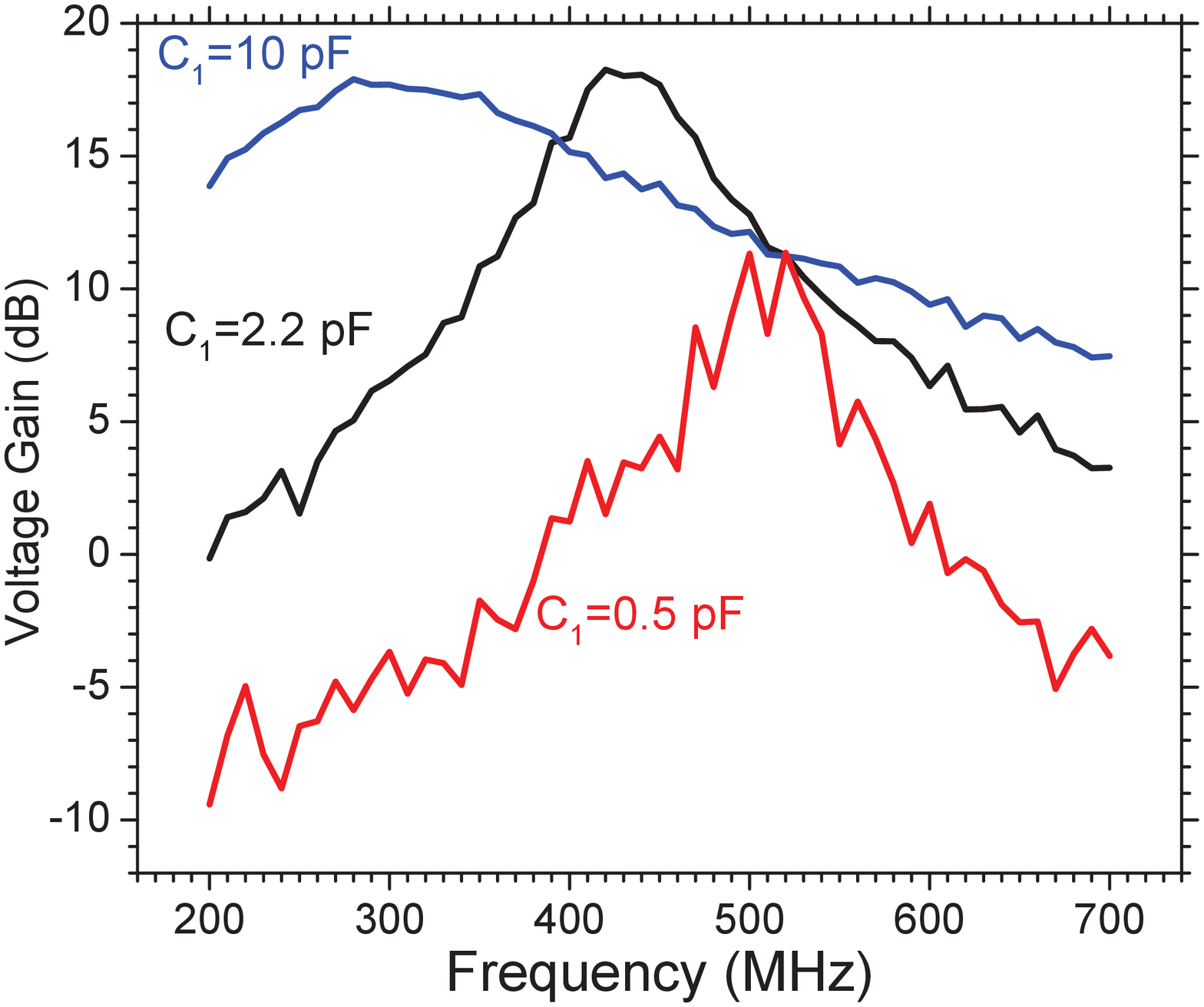}}
\vspace{-3mm} \caption{Gain of the MSA with the noisy SQUID.
$t_i=10^6t_0$; $V_i=0.1$ $\mu$V; $N=10$; $\alpha=0.001$; $T=100$ mK.
The other parameters are the same as in Fig.~\ref{reffig:gain}(a).}
\label{reffig:SN}
\end{figure}

We now show that this negative noise temperature is related to the
method of its calculation. The reason for the negative noise temperatures is in the different
methods of calculation of the noise spectral density and the gain of
the amplifier. There are three resistors that act as sources of
noise: two resistors, $R_J$, in the SQUID and the resistor, $R_1$,
in the input circuit. All three resistances contribute to the noise
spectral density, $S_V$. The noise current in the SQUID, besides
contributing to noise spectral density, changes the SQUID's
parameters, including the frequency at which the gain is maximum. At
one moment the SQUID is tuned to one frequency and at the next
moment it is tuned to another frequency. The noise voltage generated
in the resistor, $R_1$, is amplified by the detuned SQUID. On the
other hand, the signal is amplified by the SQUID tuned to a definite
frequency because there are no noise currents in the resistors,
$R_J$, of the SQUID. Consequently the amplification of the signal is
greater than the amplification of the noise. Besides, the noisy
SQUID shifts the frequency at which the amplification is maximum. In
Fig.~\ref{reffig:gain}(a) the maximum is at $f_{max}=450$ MHz, while
in Fig.~\ref{reffig:Sv} the maximum is at $510$ MHz.

\begin{figure}
\centerline{\hspace{-2mm}\includegraphics[width=8cm,height=6cm,angle=0,clip=]{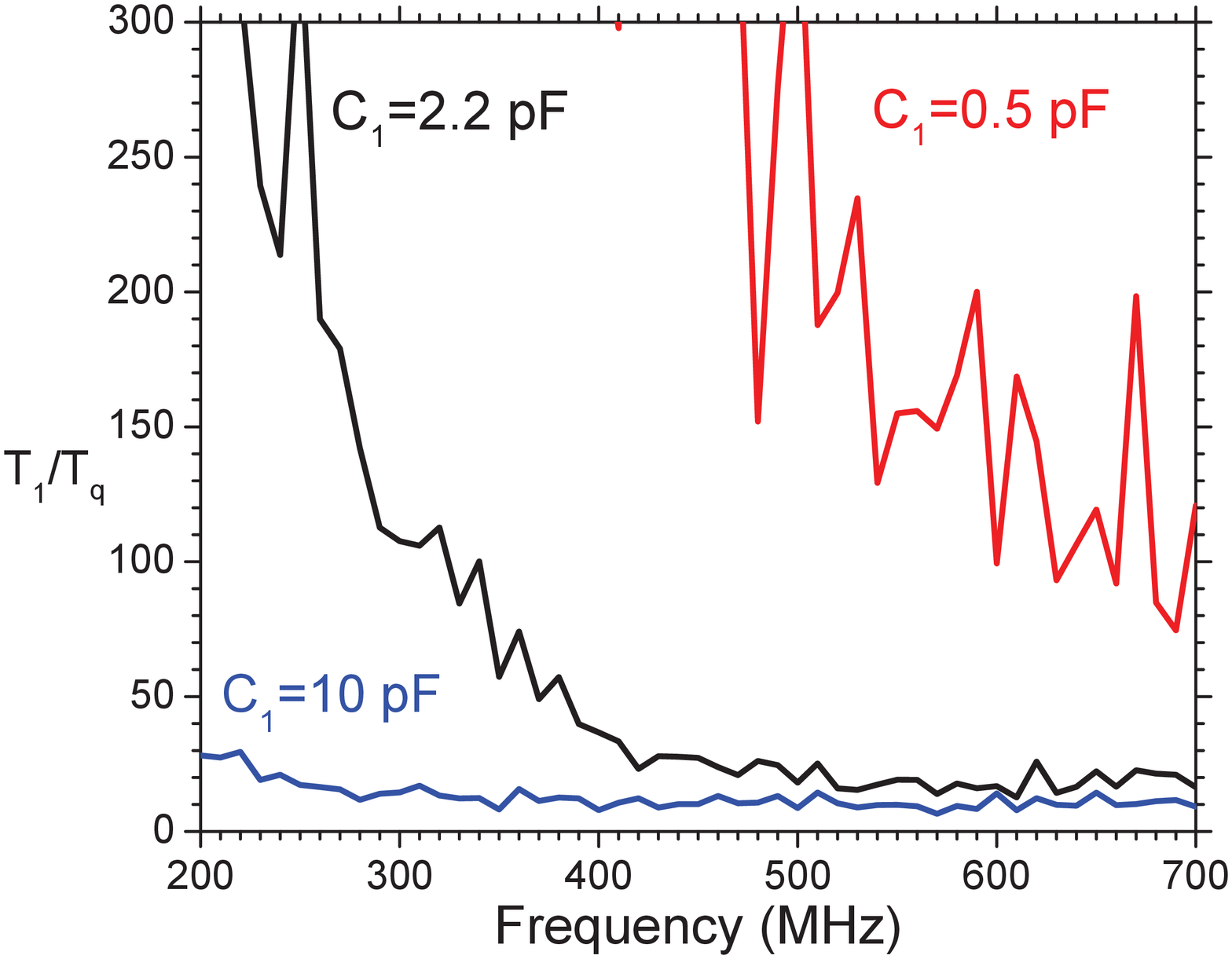}
\hspace{-2mm}\includegraphics[width=8cm,height=6cm,angle=0,clip=]{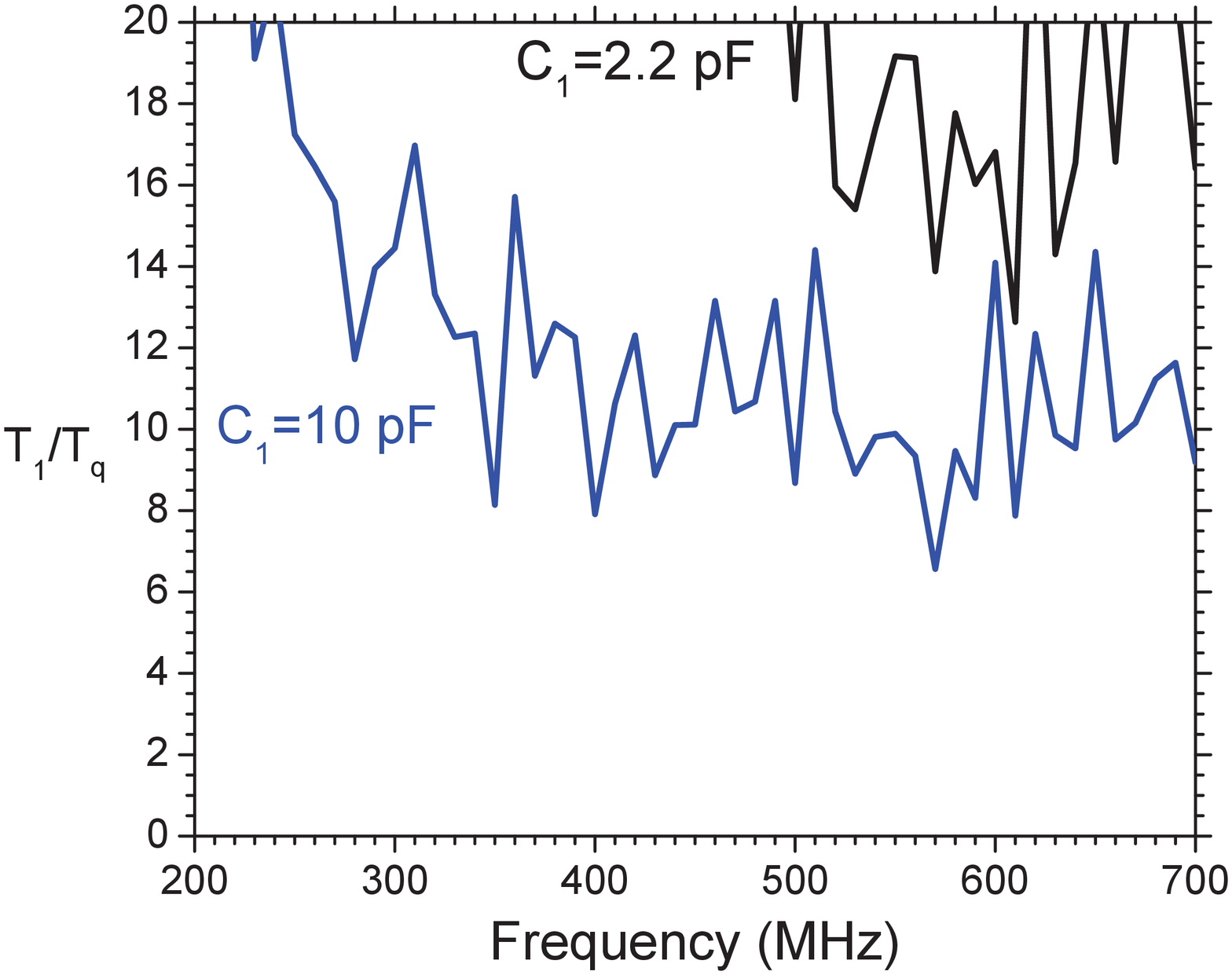}}
\vspace{-3mm} \caption{Noise temperature of the MSA with the noisy SQUID
in two different scales. The parameters are the same as in Fig.~\ref{reffig:SN}.}
\label{reffig:SNTC}
\end{figure}

\begin{figure}
\centerline{\includegraphics[width=9cm,height=7cm,angle=0,clip=]{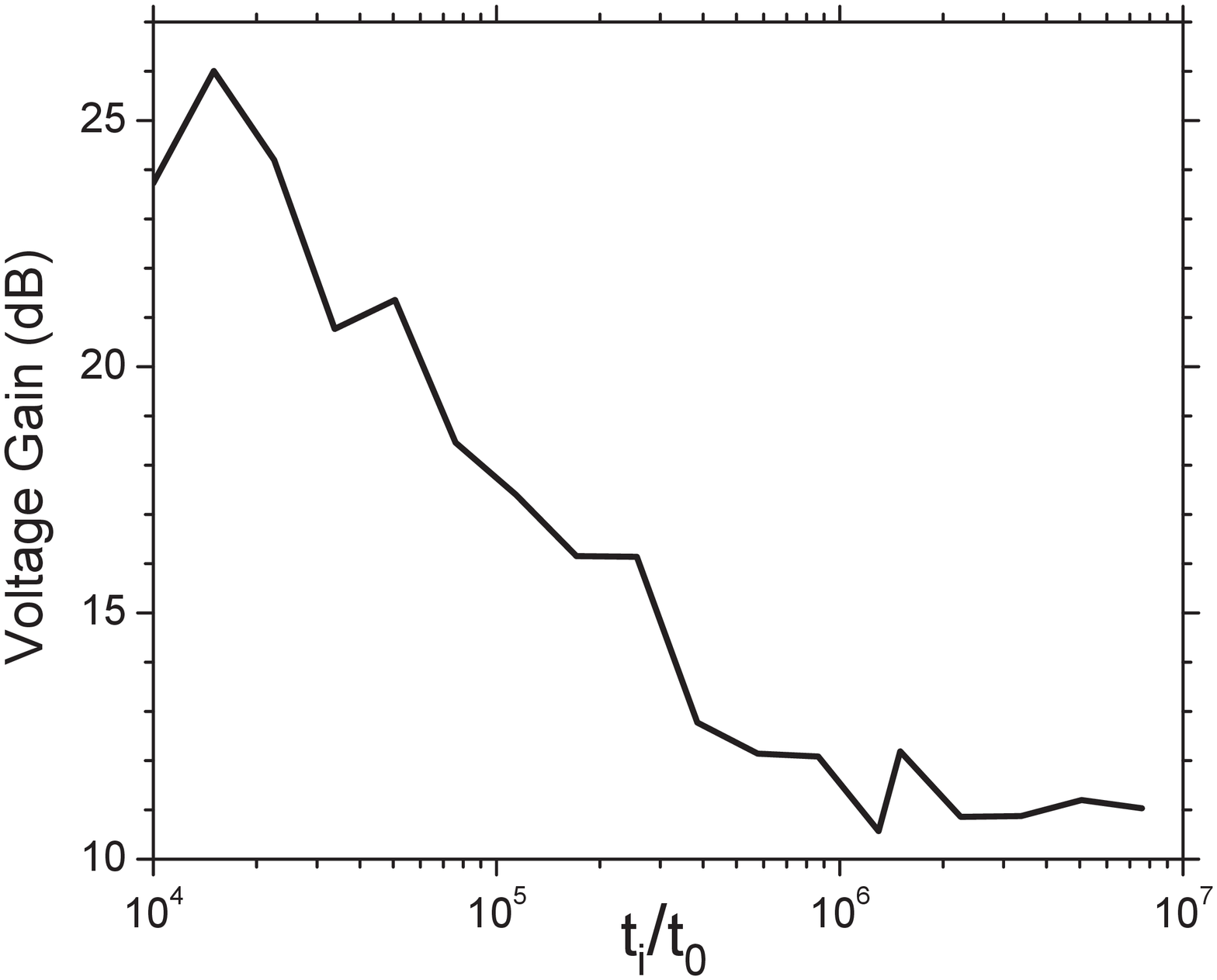}}
\vspace{-3mm}
\caption{Gain of the MSA with the noisy SQUID calculated using different integration time, $t_i$.
The contribution to the gain from the noise on the SQUID decreases as the integration time, $t_i$,
increases. For $t_i/t_0\ge 10^6$ the gain is mostly independent of $t_i$ which indicates that only
contribution from the signal remains. $C_1=0.5$ pF; the other parameters are the same as in
Fig.~\ref{reffig:SN}.}
\label{reffig:SNGtime}
\end{figure}

By this argument, we calculated the gain (Fig.~\ref{reffig:SN}) and
the noise temperature (Fig.~\ref{reffig:SNTC}) with noise on the
resistors, $R_J$, and with the input signal, $V_i\ne 0$. 
The calculated gain in Fig.~\ref{reffig:SN} is qualitatively similar to that measured experimentally
in Ref.~\cite{kinion}. If the time
of integration is sufficiently long, the contribution of the noise
to the gain is minimized, so only the contribution from the
amplified signal remains (see Fig.~\ref{reffig:SNGtime}). In this
situation both the voltage spectral density and the gain are calculated for
the same system with the noisy (detuned) SQUID.
Note that a significant asymmetry of the spectrum for $C_1 = 2.2$ pF 
and 10 pF  in Figs. 8 and 10 appears because we plot the ratio $T_1/T_q$, 
where the quantum temperature $T_q$ is proportional to the frequency $f$.  

We demonstrated two methods of calculating gain and noise
temperature of the MSA. The noise on the shunting resistors of the
SQUID reduces  gain of the amplifier if one compares
Figs.~\ref{reffig:gain}(a) (no noise) with Fig.~\ref{reffig:SN}
(with noise). The reduction is large for small capacitance,
$C_1=0.5$ pF, while the gain for $C_1=10$ pF is mostly not affected
by the noise in the SQUID. The gain calculated in the previous sections of this paper is
actually the gain at zero temperature.

In summary, we have simulated the dynamics of the microstrip-SQUID
amplifier in both the linear and nonlinear regimes and 
studied the dependence of the voltage gain and noise on the
parameters of the amplifier. We have shown that the voltage gain 
cannot exceed the critical value $G_{max}$ given by the formula (\ref{Gmax}). 
This value is inversely proportional to the input voltage. It is
shown that the gain decreases as the device temperature increases.
Finally, we have shown that the spectrum of the voltage gain depends 
significantly on the level of the Johnson noise in the SQUID resistors. 
This effect must be taken into account for correct calculation of the 
amplifier noise temperature. 
The next important step should be the optimization of the gain and noise 
temperature with respect to the amplifier's parameters. 

\section*{Acknowledgements}
This work was carried out under the auspices of the National Nuclear
Security Administration of the U.S. Department of Energy at Los
Alamos National Laboratory under Contract No. DE-AC52-06NA25396 and
by Lawrence Livermore National Laboratory under Contract DE-AC52-
07NA27344. This research was funded by the Office of the Director of
National Intelligence (ODNI), Intelligence Advanced Research
Projects Activity (IARPA).  All statements of fact, opinion or
conclusions contained herein are those of the authors and should not
be construed as representing the official views or policies of
IARPA, the ODNI, or the U.S. Government.

{}
\end{document}